\documentclass[manuscript]{aastex}

\shorttitle{2D model for the evolution of coronal plasmas}
\shortauthors{L\'opez Fuentes \& Klimchuk}

\begin{document}

\title{2D cellular automaton model for the evolution of active region coronal plasmas}

\author{Marcelo L\'opez Fuentes\altaffilmark{1}}
\affil{Instituto de Astronom\'{\i}a y F\'{\i}sica del Espacio, CONICET-UBA, CC. 67, Suc. 28, 1428 Buenos Aires, Argentina}
\email{lopezf@iafe.uba.ar}

\and

\author{James A. Klimchuk}%\altaffilmark{2}}
\affil{NASA Goddard Space Flight Center, Code 671, Greenbelt, MD  20771, USA}

\altaffiltext{1}{Member of the Carrera del Investigador Cient\'{\i}fico, Consejo Nacional de Investigaciones Cient\'{\i}ficas y T\'ecnicas (CONICET), Argentina}

\begin{abstract}
We study a 2D cellular automaton (CA) model for the evolution of coronal loop plasmas. The model is based on the idea that coronal loops are made of elementary magnetic strands that are tangled and stressed by the displacement of their footpoints by photospheric motions. The magnetic stress accumulated between neighbor strands is released in sudden reconnection events or nanoflares that heat the plasma. We combine the CA model with the Enthalpy Based Thermal Evolution of Loops (EBTEL) model to compute the response of the plasma to the heating events. Using the known response of the XRT telescope on board Hinode we also obtain synthetic data. The model obeys easy to understand scaling laws relating the output (nanoflare energy, temperature, density, intensity) to the input parameters (field strength, strand length, critical misalignment angle). The nanoflares have a power-law distribution with a universal slope of -2.5, independent of the input parameters. The repetition frequency of nanoflares, expressed in terms of the plasma cooling time, increases with strand length. We discuss the implications of our results for the problem of heating and evolution of active region coronal plasmas.
\end{abstract}

\keywords{Sun: activity -- Sun: corona -- Sun: magnetic fields -- Sun: X-rays}

%---------------INTRODUCTION--------------------------------------------

\section{Introduction}
\label{intro}

One of the most persistent conundrums of Solar Physics has been, and still is, the problem of coronal heating. The difficulties arise from both the theoretical and observational sides. Observationally, it is very difficult to explain with a single scenario the diverse set of observations obtained in different wavelengths. The first X-ray observations in the decade of 1970 suggested that coronal loops were in quasi-static equilibrium and that a steady or quasi-steady heating process was balanced by radiative and conductive losses (Rosner et al. 1978). However, the situation changed as soon as ultraviolet observations from space became available, especially when instruments like the Transition Region and Coronal Explorer (TRACE, Handy et al. 1999) began to produce higher resolution and cadence data. The evolution of many EUV loops was too dynamic and intermittent to be consistent with quasi-static evolutions. Furthermore, density determinations using EUV instruments showed that loops are too dense given their temperature and length and had scales heights that are too large to be quasi-static (Aschwanden et al. 2001, Winebarger et al. 2003). It was proposed then that EUV loops could be heated by impulsive mechanisms. Thermal evolution models based in this premise were successful in explaining many of the observed physical conditions and evolutions (see e.g., Klimchuk 2006, 2009; Reale 2010, and references therein). More recently it has been shown that impulsive heating plays an important role in the diffuse component of the corona as well (Viall \& Klimchuk 2011, 2012, 2013; Bradshaw et al. 2012, Warren et al. 2012, Schmelz \& Pathak 2012). The question is now if it is possible to understand X-ray loops, EUV loops, and the diffuse emission as part of the same phenomenon or if they are caused by completely different mechanisms. The problem has of course many different aspects: the emission evolution, geometry, location within active regions, physical conditions of the plasma, etcetera, and all of these have to be considered to provide an explanation. One proposed possibility is that all of the corona is heated by impulsive short duration events, but the rate of repetition is high at some locations (e.g., X-ray loops) and indiscernible from a continuous source (Warren et al. 2010).

From the theoretical side, and in particular regarding the heating mechanism itself, several models have been proposed. They can be grossly classified in two types:  models based on the dissipation of waves and models based on the dissipation of magnetic stresses (see the reviews by Klimchuk 2006, Reale 2010, Parnell \& De Moortel 2012). Both types can produce impulsive heating, but the best known idea is from the second category and came from Parker (1988). He proposed that loops are made of elementary magnetic strands that are shuffled and tangled by photospheric motions. Current sheets form at the boundaries between the strands, and energy is released by magnetic reconnection. Parker estimated that the energy content of a single impulsive event is roughly one-billionth of a large flare, so he coined the term ``nanoflare." We now use this term generically to describe any small spatial scale impulsive event, irrespective of the physical cause. Parker's mechanism, the basis of our investigation, could provide the impulsive events invoked in the previous paragraph to explain EUV loops, X-ray loops, and diffuse emission within the same phenomenological framework.

Several studies over the years analyzed different aspects of the nanoflare heating problem, such as the conditions for reconnection (Parker 1983a, 1983b; Priest et al. 2002, Darlburg et al. 2005), the thermal evolution of the plasma (Cargill 1994, Cargill \& Klimchuk 2004) and the observed coronal emission (Warren et al. 2010, L\'opez Fuentes et al. 2007). Other relevant issues are the mechanism by which footpoint motions translate into magnetic stress, the geometrical and temporal characteristics of the interaction between strands and the energy distribution of the produced nanoflares. A series of models based on the concept of Self Organized Criticality (SOC, Bak et al. 1988) have been developed during the last 20 years to analyze this part of the problem (see e.g., Lu \& Hamilton 1991, Lu 1995, Longcope \& Noonan 2000, Morales \& Charbonneau 2008). The idea of this kind of approach is to use simple sets of rules for the injection of energy simulating the effect of footpoint motions, the interaction between magnetic strands, and the existence of a magnetic stress threshold beyond which energy release occurs (pedagogical reviews on the subject can be found in Charbonneau et al. 2001, and the book by Aschwanden 2011).

In a recent paper we presented a simple pseudo-1D cellular automaton model to explain the observed evolution of soft X-ray loops (Lopez Fuentes \& Klimchuk 2010). Here we develop a more sophisticated model based on a similar aproach. Simulating the motions of magnetic strand footpoints in a 2D array we establish a series of rules for the interaction between strands and critical conditions for the magnetic stress relaxation through reconnection. The consequent energy release takes the form of short duration events whose output heats the plasma in the strands. To simulate the plasma response to these events we use the EBTEL code (Enthalpy Based Thermal Evolution of Loops, Klimchuk et al. 2008, Cargill et al. 2012). We analyze the output of the model by studying its dependence on the model parameters, the presence of power-laws in the energy distribution and the temporal properties of the produced nanoflares. In a second paper we will compare the results of the model with observed X-ray and EUV loops.

The paper is organized as follows. In Section~\ref{model} we provide a detailed description of the model, the implementation of EBTEL to compute the plasma response and the obtainment of synthetic observations. In Section~\ref{analysis} we present and analyze the results and discuss some of their implications and conclude in Section~\ref{conclusions}.

%--------------MODEL DESCRIPTION----------------------------------------

\section{Description of the model}
\label{model}

\begin{figure*}[t]
\centering
\hspace{0.cm}
\includegraphics[bb=14 14 888 448,width=16.cm]{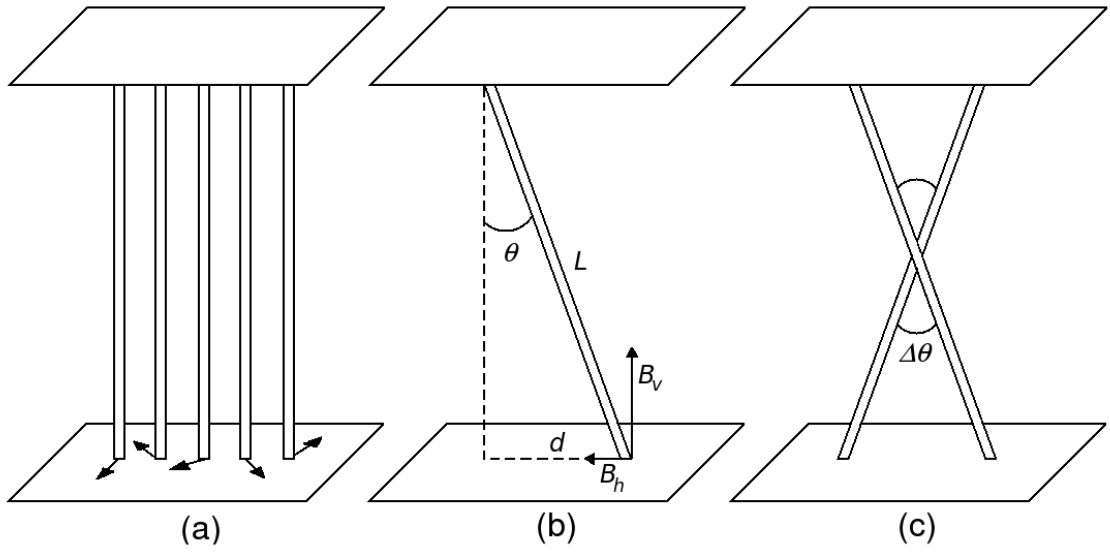}
       \caption{Schematic description of the model (see Section~\ref{model} for details).}
\label{cartoon}
\end{figure*}

The model studied here is a more sophisticated version of the simple 1D model presented in L\'opez Fuentes \& Klimchuk (2010). In the present version magnetic strand footpoints are represented by moving elements on a 2D grid. The idea behind the model can be explained starting from the classic view (Parker 1988) shown in Figure~\ref{cartoon}, that represents coronal magnetic flux tubes or ``strands'' as vertical structures connecting two remote sections of the photosphere (represented by horizontal planes). In the initial configuration of panel (a), all flux tubes and their corresponding axial magnetic fields are vertical. As the strand footpoints are dragged by photospheric motions, magnetic stress is induced in the configuration progressively increasing the free magnetic energy. In the geometrical representation of Figure~\ref{cartoon}, the displacement of the footpoints (the distance $d$ shown in panel (b)) is associated with an inclination angle ($\theta$) and the appearence of a horizontal component ($B_h$) of the strand magnetic field. Naming $B_v$ the vertical field, it can be seen from Figure~\ref{cartoon}, panel (b), that

\begin{equation}
B_h = B_v \tan\theta \approx B_v \frac{d}{L},
\end{equation}

\noindent where $L$ is the strand length. The approximation is accurate as long as the strand is not strongly inclined, as we expect. When sections of neighboring strands are in contact, their mutual inclination ($\Delta\theta$, see panel (c)) implies a current sheet at the interface. Dahlburg et al. (2005, 2009) have shown that the current sheet is unstable to the secondary instability whenever the misalignment exceeds a critical angle. Explosive reconnection occurs releasing magnetic energy and heating the plasma impulsively.

\begin{figure*}[t]
\centering
\hspace{0.cm}
\includegraphics[bb=14 14 580 319,width=15.cm]{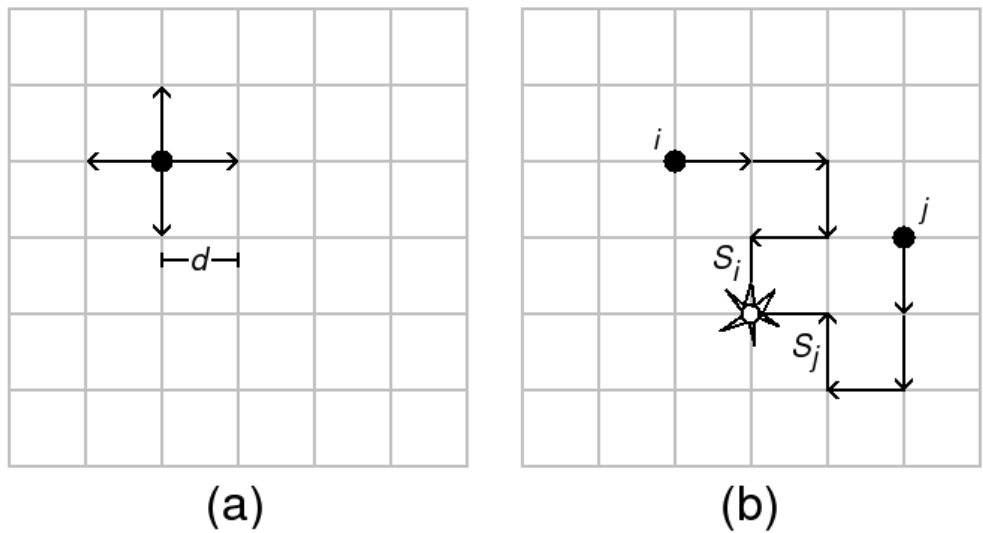}
       \caption{Description of the cellular automaton evolution (see Section~\ref{model} for details).}
\label{grid}
\end{figure*}

In our model, the above scenario is reduced to 2 dimensions in the following way. As it is shown in Figure~\ref{grid}, the strands are represented by moving elements in a plane grid. Initially, each grid node is occupied by a single strand element. To simulate the displacement of the strand footpoints by photospheric motions, at each time step each element is randomly displaced one grid step in any of 4 directions as shown in panel (a).
The boundary conditions are periodic, meaning that each time an element crosses the mesh border it reenters the grid on the opposite side. The physical justification is that to keep the density of strands constant, each element that leaves the system will on average be replaced by a similar one. The distance between grid nodes corresponds to the typical size of a convective granule ($d\sim$1000 km) and we associate the time step duration with the convective turnover time scale ($\delta t \sim$1000 sec). After a certain number of time steps, strand element $i$ would have been displaced across the grid a distance $S_i$ (see panel (b)). Following the arguments of the previous paragraph, this displacement produces a horizontal component of the magnetic field that can be approximated as

\begin{equation}
B_h^i  \approx \frac{B_v}{L} S_i.
\end{equation}

\noindent As will become more clear in the following paragraphs, $S_i$ should not be regarded as a distance on the grid \textit{per se}, but as a length variable associated with the inclination of the strand. Consider now the situation from panel (b), in which strand element $i$ encounters $j$, that has its own displacement variable $S_j$. In the model, we relate this encounter to the situation described in Figure~\ref{cartoon}, panel (c). By similar arguments to those used for Equations~1 and~2, the misalignment angle can be approximated as

\begin{equation}
\label{delta_theta}
\Delta \theta \approx \tan^{-1} \left(\frac{S_i+S_j}{L}\right).
\end{equation}

Note that the grid in Figure~\ref{grid} should {\it not} be thought of as the photosphere. It is simply a conceptual way of treating the chance (coronal) interaction of strands in a tangled magnetic field. We assume that the motions of the strands are uncorrelated. If this assumption were violated, the misalignment angle would be less than indicated in Equation~3.

Assuming the existence of a critical condition for the energy release, each time two strand elements occupy the same node we compare the inclination angle, $\Delta \theta$, with a given critical value $\theta_c$. If $\Delta \theta > \theta_c$, we simulate reconnection by relaxing the magnetic stress between the two strands through the transformation of variables $S_i$ and $S_j$, as follows:

\begin{eqnarray}
\label{relaxation1}
 S_i' = \alpha (S_i-d) + (1-\alpha) (S_j-d),\\
\label{relaxation2}
 S_j' = (1-\alpha) (S_i-d) + \alpha (S_j-d),
\end{eqnarray}

\noindent where $\alpha$ is a random number between 0.2 and 0.8, that accounts for the fact that the reconnection between strands $i$ and $j$ is not symmetric in general. The strands essentially exchange legs, and these segments can be long or short depending on how high in the corona (at what location between the planes in  Figure~\ref{cartoon}) the reconnection takes place. The transformation given by Equations~\ref{relaxation1} and~\ref{relaxation2} implies that each time a strand reconnects, its ``displacement'' variable $S_i$ is on average reduced by an amount $d$. The choice is not arbitrary, since after reconnection the expected entanglement with other strands will prevent the reconnecting pair from relaxing much more than the typical separation between strands ($d$). This average decrease coincides with the increase received at each time step due to the footpoint displacement process, which is therefore the typical approximate excess over the critical condition. The relaxation returns the strands to a state close to, and not far below, the critical level of stress.

 With each iteration, we move each strand to an adjacent position on the grid and increase $S_i$ by an amount $d$. If there is an encounter with one or more other strands at that same position, we decrease $S_i$ as described above. However, since we do {\it not} move the strand back to its old position, after a few transformations, $S_i$ rapidly departs from the length of the path on the grid. (It is smaller than the path length.) As we stated above, $S_i$ should not be regarded as a distance, but as a variable associated with the strand inclination in the sense given by Equation~\ref{delta_theta}. We emphasize that the grid does not represent the photosphere, or a physical surface at any height. Instead, it should be regarded as a convenient 2D version of the 3D scenario illustrated in Figure~\ref{cartoon}. Its purpose is to model the random aspect of strand pair interactions. At a given iteration, a strand may reconnect with one, several (rarely more than 3), or no other strands, depending on how many occupy the same node. This realistically captures the property that strands are in contact with multiple neighbors in a tangled 3D magnetic configuration.

 Operationally, it is important to ``order'' the testing and reconnection of the strands in the system. If the strands are tested and reconnected in order, for instance, starting with strand 1, then 2, then 3, and so on, most of the nanoflares will occur only in the first strands. There can be strands that suffer less nanoflares because their interacting partners already reconnected with some of the first strands and relaxed. In the actual code, we solve this by identifying first all the critical pairs. We then pick one randomly, and once this first pair has been reconnected we retest for all the remaining critical pairs again, then we randomly pick and reconnect another one, and so on until no critical pairs remain. At that point the program advances one timestep. With this scheme if one strand of a pair remains critical after its first reconnection it will eventually be reconnected a second time in the same step. Of course, no more than one or two reconnections per strand are expected.

 We remind the reader that the idea of the present approach is to reproduce in a physically reasonable, but still geometrically simple way, the scenario of footpoint motion producing magnetic stress between interacting strands that leads to energy release in the form of short duration reconnection events. The simple displacement of points in a 2D grid is meant to reproduce the random nature of the strand mixing in the corona. The main motivation for the model simplicity is to save computing time, so a very large number of strands and nanoflares can be easily handled within runs of durations from minutes to tens of minutes. The combination with a fast code like EBTEL makes this choice even more convenient, so diverse sets of parameter combinations can be tested in manageable times. Such a study is not possible with a full 3D MHD treatment.

Noting that $S_i'+S_j'=S_i+S_j-2d$, the transformation implies, through Equation 3, a relaxation of the mutual inclination of the strands. Considering the association between $B_h^i$ and $S_i$ (correspondingly $j$) given by Equation 2, it can be easily shown that the above transformation implies a difference of magnetic energy density,

\begin{equation}
\label{energy_eq}
\Delta e_{ij} = \frac{B_v^2}{8\pi L^2}\left[\left(S_i'^2+S_j'^2\right)
- \left(S_i^2+S_j^2\right)\right].
\end{equation}

\noindent We consider that each of these energy releases heats the plasma in the strands in the form of nanoflares. Since we do not expect the heat to be equally distributed between the strands, we include a second random number $\beta$ (0.25 $<\beta<$ 0.75), so strand $i$ receives an energy $e_i = -\beta \Delta e_{ij}$ and strand $j$ receives $e_j = -(1-\beta) \Delta e_{ij}$. At each time step we record the total energy per unit volume received by each strand due to the interaction with all of its critical partners. As previously discussed, depending on the initial $S_i$ and $S_j$, the transformation from Equation~\ref{relaxation1} and~\ref{relaxation2} could be insufficient to fully remove the critical condition ($\Delta \theta > \theta_c$). Therefore, the procedure is repeated for all interacting strands until no critical pairs remain. Once this is accomplished, the system is allowed to evolve to the next time step. The process repeats for as many time steps as wished. The longer the evolution of the system, the larger the nanoflare statistics produced.

The main input parameters of the model are the strand length ($L$), the vertical magnetic field ($B_v$), the critical misalignment angle ($\theta_c$) and the number of strands ($N$). We set these parameters to reasonable solar values and vary them, each at a time, to investigate the effect that they have on the model's output. As explained above, parameters such as the photospheric displacement of the footpoints ($d$) and the time step duration ($\delta t$) are set at 1000 km and 1000 sec respectively, for all runs. The output of the model is the recorded series of nanoflares suffered by each strand at each time step.

One of the main motivations of the present model is the comparison with actual coronal observations. To accomplish this we need to produce synthetic observations from the model's output described above. To simulate the response of the plasma contained in each strand to the heating produced by the nanoflares, we use the 0D hydrodynamic code EBTEL (see Klimchuk et al. 2008, Cargill et al. 2012). Modeling each nanoflare as a triangular function of a given duration, the EBTEL code provides us the temperature and density evolution of the plasma, which we combine with the known response of particular coronal instruments to obtain simulated strand light-curves. Since an observation will include many unresolved strands, due to both the optically thin nature of the corona and the finite resolution of real instruments, we add the emission of all strands to obtain the evolution of the total intensity per pixel registered by the instrument. It is worth noting here that the cellular automaton model and the EBTEL model work on different temporal scales. While CA time steps last 1000 s, individual nanoflares are set to several hundred seconds, and EBTEL time steps last one second.

In this work we simulate observations obtained with the X-ray Telescope (XRT, see Golub et al. 2007) on board Hinode (Kosugi et al. 2007), to study the effect of different input parameter combinations on the modeled emission. In an accompanying observational paper we compare the model with coronal data in different wavelengths obtained with the Hinode/XRT and SDO/AIA (Pesnell et al. 2012, Lemen et al. 2012) instruments.

%---------------------MODEL RESULTS------------------------------------------

\section{Analysis of the results}
\label{analysis}

\subsection{Scaling relations}
\label{scalings}

\begin{figure*}[t]
\centering
\hspace{0.cm}
\includegraphics[bb=120 30 470 750,width=9.cm]{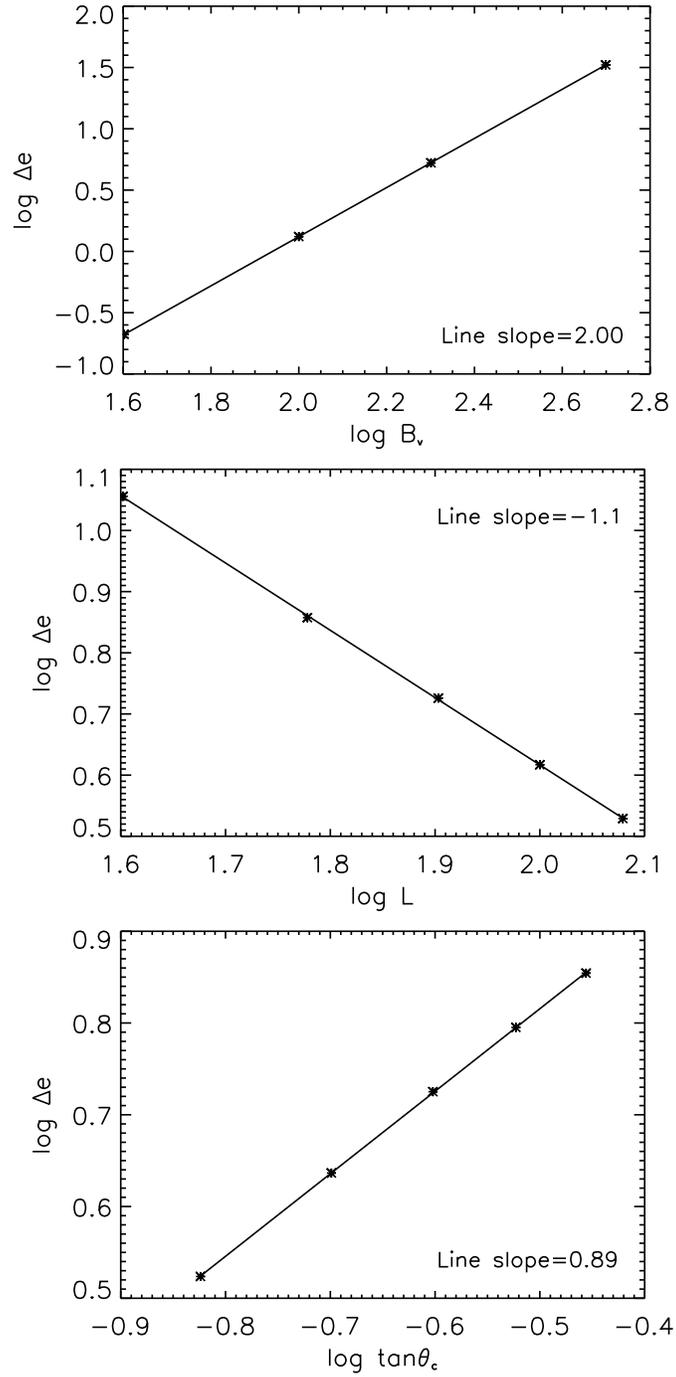}
       \caption{Log-log plots of the nanoflare energy versus relevant parameters of the model. The lines correspond to linear regressions of the plotted points. Line slopes from the regressions are provided in the corresponding panels (see Section~\ref{scalings} for details).}
\label{energy}
\end{figure*}

\begin{figure*}[t]
\centering
\hspace{0.cm}
\includegraphics[bb=55 155 580 690,width=16.cm]{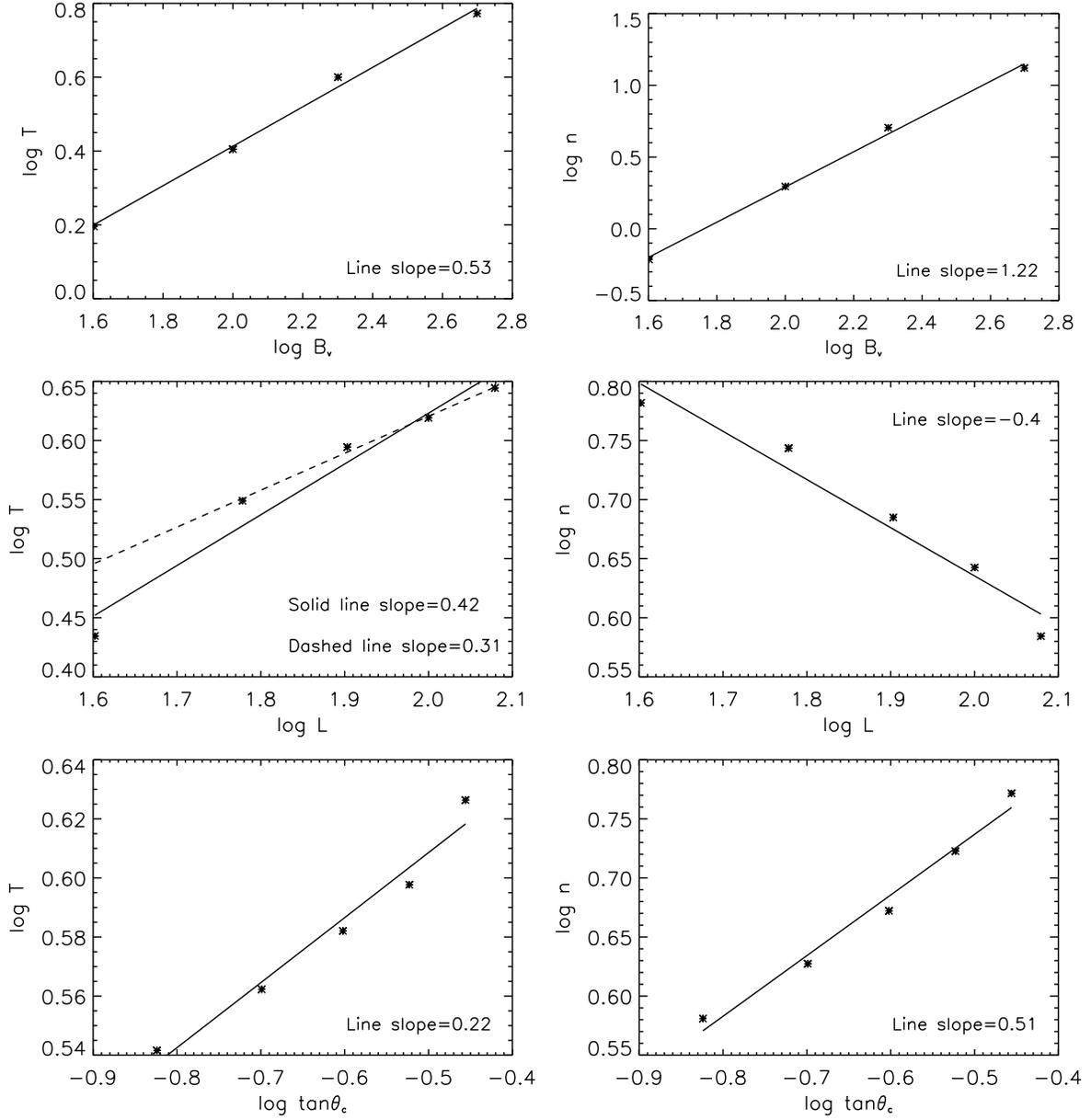}
       \caption{Idem Figure~\ref{energy} for the plasma temperature and density obtained with the EBTEL code (see Section~\ref{scalings} for details).}
\label{plasma}
\end{figure*}

\begin{figure*}[t]
\centering
\hspace{0.cm}
\includegraphics[bb=130 25 470 755,width=9.cm]{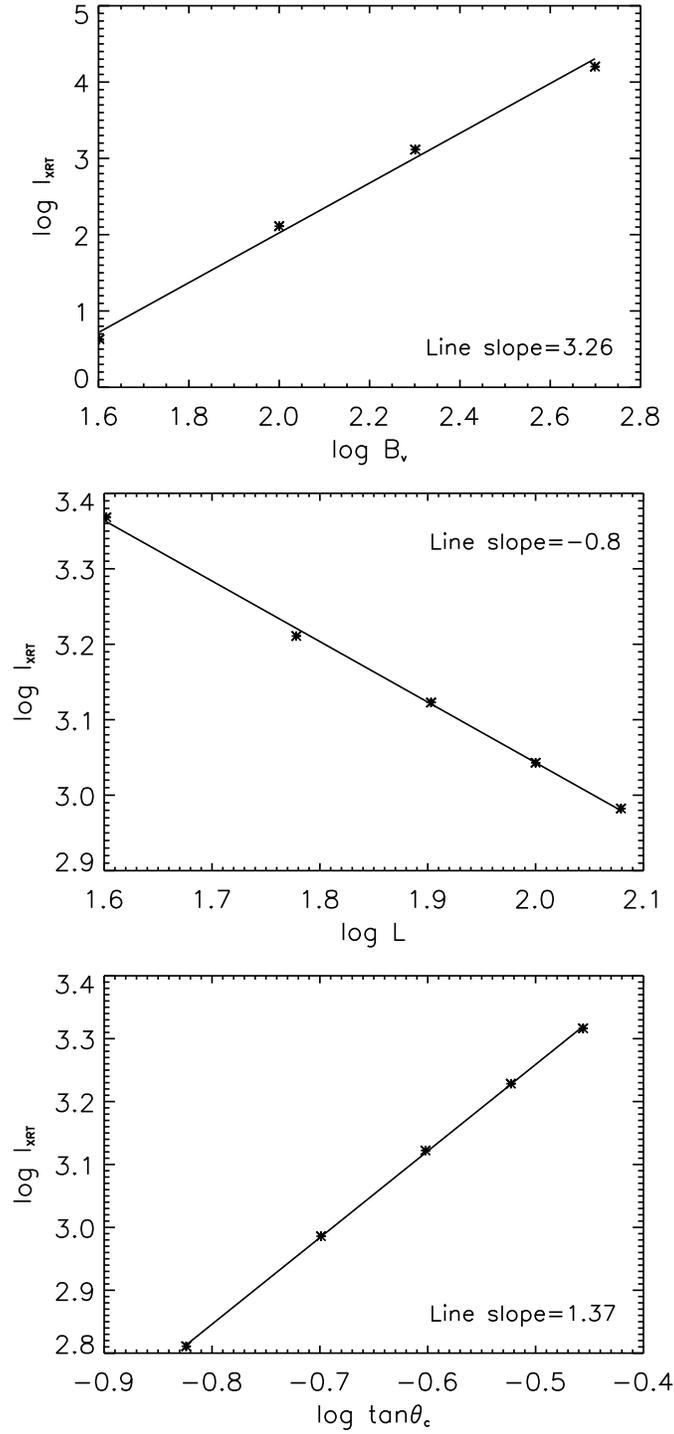}
       \caption{Idem Figures~\ref{energy} and~\ref{plasma} for the simulated XRT intensity (see description in Section~\ref{scalings}).}
\label{xint}
\end{figure*}

To study the effect of the main input parameters on the model's output, we choose a set of values appropriate for the Sun, and we vary them one at a time leaving the rest fixed. In Table~\ref{parameters} we list the different parameters indicating in bold face the set of values that remain fixed during individual variations. As we show below, $B_v$, $L$ and $\tan \theta_c$, are the parameters that have the largest effect on the model's output. We focus on how they affect the mean energy of the nanoflares produced by the model, the plasma density and temperature computed with EBTEL, and the synthetic XRT intensity obtained as described in Section~\ref{model}. For all the EBTEL simulations we use a time resolution of 1 s. Figures~\ref{energy}, \ref{plasma} and \ref{xint}, show plots of the corresponding mean values as a function of the different input parameters tested. In all cases the log-log plots indicate the clear presence of power scalings. The lines correspond to linear regressions of the data and their slopes are indicated in the panels. In what follows we analyze the physical meaning of the scaling relations.

Let us begin with the mean nanoflare energy density ($\Delta e$) that heats the strands (see Figure~\ref{energy}). If we consider that, on average, the ``distance'' variables of two interacting strands are similar ($S_i \approx S_j$), and that after reconnection each of them is roughly reduced by a distance $d$, we can approximate the energy expression from Equation~\ref{energy_eq} as

\begin{equation}
\Delta e \approx \frac{B_v^2}{4\pi L^2}\left[\left(S_i-d\right)^2
- S_i^2\right] \approx \frac{B_v^2 d}{4\pi L^2}\left(-2S_i+d\right).
\end{equation}

\noindent Since in the fully developed regime $2S_i>>d$,

\begin{equation}
|\Delta e| \approx \frac{B_v^2 d}{2\pi L^2}S_i.
\end{equation}

\noindent It is easy to see from Equation~\ref{delta_theta} that $S_i \approx (L/2) \tan \theta_c$. Therefore, we can write the following scaling relation for the mean nanoflare energy:

\begin{equation}
\label{e_scaling}
|\Delta e| \propto \frac{B_v^2}{L}\tan \theta_c.
\end{equation}

\noindent The above expression explains the scalings obtained from the linear fittings shown in Figure~\ref{energy}.

The explanation of the scalings shown in Figure~\ref{plasma} requires the analysis of approximate relations for the plasma response to the heating. As discussed by Vesecky et al. (1979), in quasi-static conditions, the average heating rate ($Q$), conductive loss rate, and radiative loss rate are all comparable:

\begin{equation}
\label{balance}
Q~\approx \frac{2}{7} \kappa_0 \frac{T^{7/2}}{L^2} \approx n^2 \Lambda(T).
\end{equation}

\noindent The second term is an approximation for the divergence of the heat flux. Here, $\kappa_0$ is the coefficient of thermal conduction, $T$ and $n$ are the temperature and density of the plasma, and $\Lambda(T)$ is the radiative loss function. One might imagine that the above relation is only valid in the high-frequency nanoflare regime that resembles a quasi-steady state. However, it can also be reasonable if low-frequency nanoflares dominate. Cooling after a nanoflare occurs in three stages. Thermal conduction dominates early, radiation dominates late, and the two are comparable at intermediate times. This middle stage lasts the longest, so median values of temperature and density will be representative of this stage and therefore of quasi-static conditions. Also, depending on the temperature sensitivity range of the observing instrument and on the nanoflare energy, the strands may only be visible when they are in this middle stage of cooling. The radiation loss function is usually expressed as

\begin{equation}
\label{lambda}
\Lambda(T) \approx \Lambda_0 T^b,
\end{equation}

\noindent where $\Lambda_0$ and $b$ are constants within given temperature intervals. From Rosner et al. (1978), for the temperature range of interest here, $b$ = -0.5. From Equation~\ref{balance} we can obtain the following relation between the plasma temperature and the heating rate:

\begin{equation}
\label{t_q_scaling}
T \propto Q^{2/7} L^{4/7}.
\end{equation}

\noindent In the present model, the heating rate on each strand is the average nanoflare energy ($\Delta e$) divided by the mean waiting time between events. As discussed in Section~\ref{model}, due to the transformations from Equations~\ref{relaxation1} and~\ref{relaxation2}, a strand would need on average a single time step to recover the critical condition. It is expected then, that each strand experiences an average of one nanoflare per time step, regardless of the model parameters. We confirm this by directly counting, from the model output, the number of nanoflares that occur in each strand at each time step. Then, we conclude that $Q \propto \Delta e$. Replacing this in Equation~\ref{t_q_scaling} and using the scaling from Equation~\ref{e_scaling}, we finally have for the temperature:

\begin{equation}
\label{t_equation}
T \propto B_v^{0.57} L^{0.28} \left(\tan \theta_c \right)^{0.28}.
\end{equation}

\noindent Here, we write the exponents in decimal form for an easier comparison with the line slopes provided in the corresponding panels of Figure~\ref{plasma}. Notice that the slope of the solid line in the $T$($L$) plot (second row, left panel), which includes all the $L$ values, slightly departs from the above expected exponent. The reason is that the quasi-static approximation assumed to obtain Equation~\ref{t_equation} is less applicable for shorter loops, in which the evolution produced by the impulsive heating is much more dynamic. To confirm this, we make a second line fitting (dashed line) that does not include the $L$= 40 Mm case. We then obtain a slope (0.31) that is much closer to the expected value (0.28).

Using Equations~\ref{balance} and~\ref{lambda} it can be easily shown that
$n \propto T^2/L$. Replacing $T$ from Equation~\ref{t_equation} in the previous expression, we obtain for the density:

\begin{equation}
\label{n_equation}
n \propto B_v^{1.14} L^{-0.43} \left(\tan \theta_c \right)^{0.57}.
\end{equation}

\noindent The line slopes provided in the right panels of Figure~\ref{plasma} are consistent with the expected exponents. Once again, small departures are due to the limitations of the quasi-static approximation used.

The previous relations can be used to explain the intensity scalings obtained from the line fittings shown in Figure~\ref{xint}. The simulated XRT intensities whose mean values are plotted in the figure are obtained using the equation

\begin{equation}
I_{XRT} = n^2 S(T),
\end{equation}

\noindent where $n$ and $T$ are the density and temperature computed with EBTEL, and $S(T)$ is the known response of the XRT instrument in the Al-poly filter position. Fitting the instrument response with a function of the form $S(T) = S_0 T^a$, we find that for the temperature range of interest here, $a$=1.92. Using the $n$ and $T$ scalings from Equations~\ref{t_equation} and~\ref{n_equation} and the previous relations we finally obtain

\begin{equation}
I_{XRT} \propto B_v^{3.37} L^{-0.33} \left(\tan \theta_c \right)^{1.67}.
\end{equation}

\noindent The comparison between the exponents of the previous relation and the computed slopes in Figure~\ref{xint}, show that although the $B_v$ and $\tan \theta_c$ exponents are in good agreement, the one corresponding to $L$ is not. This is possibly due to the several approximations used to obtain the above expression.

\begin{figure*}[t]
\centering
\hspace{0.cm}
\includegraphics[bb=90 370 550 700,width=13.cm]{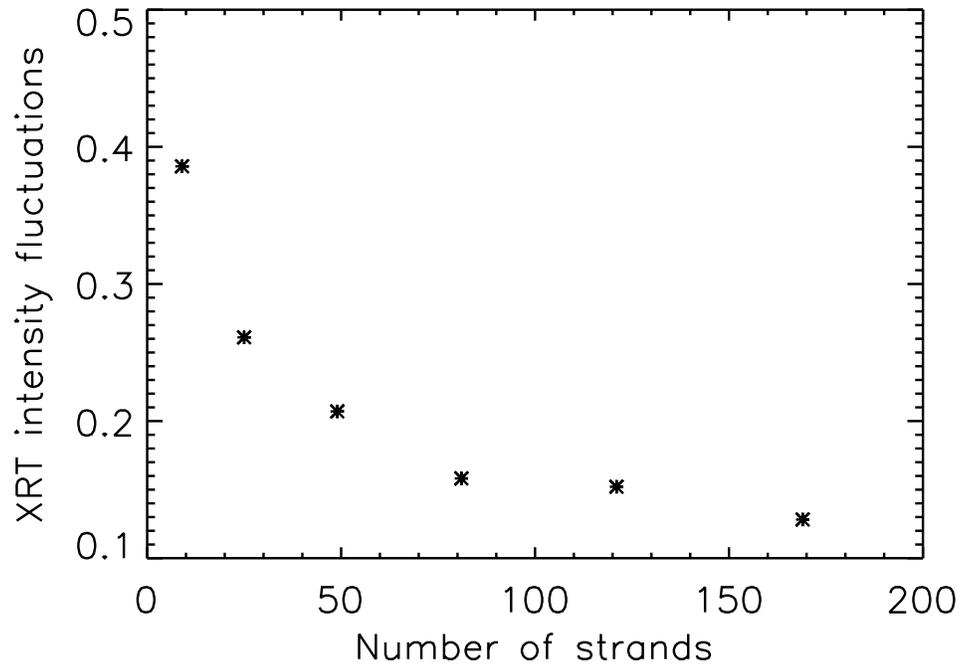}
       \caption{Amplitude of the fluctuations of the XRT synthetic intensity as a function of the number of strands used in the model.}
\label{nstrands}
\end{figure*}

We expect that fluctuations of the simulated XRT signal will depend on the number of strands, $N$, that are are included in the summation. The heating-cooling processes due to individual nanoflares acting in different strands produce much more variability in the intensity signal when $N$ is small than when it is large. In the latter case, the sum of many strand intensities tend to smear out the ups and downs of individual evolutions. To illustrate the effect of number of strands on the XRT synthetic intensity, in Figure~\ref{nstrands} we plot the relative amplitude of the intensity standard deviation versus $N$. As can be noticed from the figure, the relative size of the fluctuations reaches approximately 0.15 or less for $N \gtrsim 100$. As we show in the observational study, this is consistent with what is found in real observations. Note that the larger the number of strands, the smaller the strand cross section must be in order to fit into the same volume.

 XRT signal fluctuations are also affected by the nanoflare duration, $\tau$, because shorter nanoflares tend to produce a more marked heating/cooling alternation than longer events. Thus, the amplitude of the intensity fluctuations is expected to decrease with an increase of the nanoflare duration. We analyzed this in detail in L\'opez Fuentes \& Klimchuk (2010) for a broad range of $\tau$ values. In the present study though, we adopt the conservative approach to consider that the nanoflare durations are not expected to be longer than the CA time step duration ($\delta t \approx$ 1000 sec, see previous Section). We remind the reader that the plasma evolution is modeled with EBTEL using a time resolution of 1 s. For the range of $\tau$ values presented in Table~\ref{parameters} we do not find substantial variation of the intensity fluctuations.

The scaling relations found in this Section can be used as test predictions of the model to be compared with future observations.

%----------------POWER LAWS---------------------------------------------

\subsection{Nanoflare energy distribution}
\label{power_laws}

 Since the model presented here is based on cellular automata evolution, the question arises whether it shares some of the known properties of similar models that produce self organization. One of the most salient features of Self-Organized Critical models (SOC, see the review by Charbonneau et al. 2001) is the existence of power laws in the distributions of energies, durations and geometric properties of the events produced. These models have been invoked to explain the presence of power law distributions in several solar phenomena, from flare observations in different wavelengths and energy ranges to Solar Energetic Particle (SEP) events (for a full list of references see Aschwanden 2011). As part of our effort to explain the observed properties and evolution of the plasma in coronal loops, we here examine the presence of power law distributions in the nanoflare energies produced by the model.

\begin{figure*}[t]
\centering
\hspace{0.cm}
\includegraphics[bb=55 225 515 570,width=13.cm]{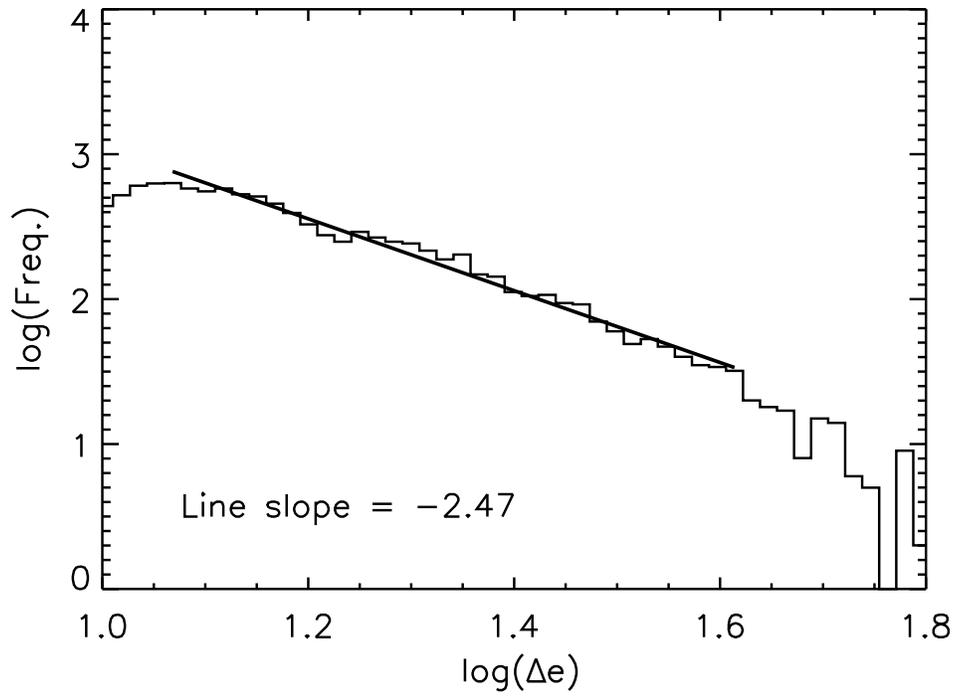}
       \caption{Power-law distribution of nanoflare energies.}
\label{power-law}
\end{figure*}

For each of the different parameter combinations in Table~\ref{parameters}, we produced a nanoflare energy distribution, i.e., log-log histogram of the number of events as a function of energy. We then identified (by eye) and fitted (by regression) the linear portion of the distribution to determine its slope. Figure~\ref{power-law} shows one of these distributions, with the linear part indicated with a thick line segment and the corresponding slope provided in the panel. We find that all of the models have approximately the same slope. There is no evidence for a dependence on the model parameters. The mean slope for all of the parameter combinations in  Table~\ref{parameters} is -2.54 and the standard deviation is 0.17. The minimum and maximum values are -2.9 and -2.14. Different runs using the same set of parameters provide slopes with the same type of distribution as for the full parameter variation. This suggests that the mean slope of approximately -2.5 is a robust feature associated with the kind process described by the model, regardless of the particular choice of parameters. The slight differences in the slopes are likely due to the binning and selection procedure and the random nature of the modeled system.

We wish to know whether the power law is valid over a wider range of energies than are represented by the parameters in Table~\ref{parameters}. For those runs, $B_v$ and $L$ were varied independently. However, since magnetic fields generally weaken with distance from their source, we expect long strands to have statistically weaker fields than short strands. Mandrini, D\'emoulin \& Klimchuk (2000) found that $B \propto L^{-0.88}$ in active regions. We also know from Equation~\ref{e_scaling} that nanoflares tend to be more energetic in strands with shorter lengths and stronger fields. Combining these two results, it is clear that nanoflare energies will vary considerably over an active region. We performed five new simulations using Mandrini et al. to relate $B_v$ and $L$:

\begin{equation}
\label{b_lock}
B_v \approx 10^4  \times L^{-0.88},
\end{equation}

\noindent where $B_v$ is expressed in Gauss and $L$ in Mm, and the constant of proportionality is courtesy of C. H. Mandrini (private communication). The other model parameters were fixed at $\tan \theta_c = 0.25$, $N = 49$ and $\tau = 200$ sec.  Table~\ref{parameters2} gives the combinations of $B_v$ and $L$ and corresponding range of nanoflare energies $\Delta E$ over which the distribution exhibits a power law with the indicated index. Notice that $\Delta E$ corresponds to the total energy of the nanoflares, not the energy density, $\Delta e$, analyzed in previous sections. The energies have been integrated over the strand volume assuming a strand cross section that is consistent with the properties of observed coronal loops. We consider that the $N$ strands of the model fill a loop of a typical observed diameter. We see that the indices are very similar even though the energy ranges are much different. The mean is -2.5 and the standard deviation is 0.17, similar to the values obtained with the full set of parameters of Table~\ref{parameters}. This provides further support for the existence of a universal power law. The fluctuations of the index are once again likely due to the errors in the determination of the linear part of the distribution. To test this we repeat several times the run with $L = 80$ Mm ($B_v = 211$ G) obtaining a mean and standard deviation which are consistent with the previous results.

Each of the numerical experiments described in the above analysis corresponds to model runs of 200 time steps. Considering that each strand is heated at an average rate of one nanoflare per time step and that $N = 49$ for all the runs, each energy distribution includes approximately 9800 nanoflares. To test the statistical robustness of the power-law indices, we performed an experiment with 50000 time steps, using $L = 80$ Mm ($B_v = 211$ G) and the rest of the parameters as indicated in the above paragraphs. In this case, the number of nanoflares in the sample increases to approximately 2.45$\times 10^6$. The index is -2.57, in agreement with the previous results.

It is interesting to analyze the origin of the energy ranges shown in the third column of Table~\ref{parameters2}. For a given set of parameters, the reconnection rules imposed by Equations~\ref{relaxation1} and~\ref{relaxation2} imply that strands are never far from their critical state. The nanoflare energies are expected to be distributed around a value proportional to $\frac{B_v^2}{L}\tan \theta_c$ (see Equation~\ref{e_scaling}), with variations due to the randomness of the strand mixing and the redistribution of strand lengths and released energies produced by the use of the random variables $\alpha$ and $\beta$. The combination of the 2D strand mixing and the random variables produce energy distributions with a prominent linear part. The constancy of the slope of the linear part for a wide range of model parameter combinations suggests the presence of a scale invariance associated with the procedure used for the strand interaction and reconnection in the model.

%----------------FREQUENCY---------------------------------------------

\subsection{Nanoflare frequency}
\label{frequency}

As discussed in the Introduction, it has been argued that nanoflares can explain a variety of observations (X-ray loops, EUV loops, active region cores, etc.) depending on whether the delay between successive events is long or short compared to a cooling time. Long delays correspond to low-frequency nanoflares, and short delays correspond to high-frequency nanoflares. With low-frequency nanoflares, the plasma has time to cool fully before the next event occurs. This results in a broad distribution of temperatures. With high-frequency nanoflares, the plasma is reheated after only a small amount of cooling, so the temperature fluctuates about a mean value. There are actually three important timescales in determining the thermodynamic properties of the plasma: the nanoflare repetition time, or interval between the start of consecutive events, the cooling time, and the nanoflare duration. If the duration is long, there may be little time for the plasma to cool even if the repetition time would suggest low-frequency heating. In this case the nanoflares are effectively high frequency. We examine the influence of all three of these timescales.

\begin{figure*}[t]
\centering
\hspace{0.cm}
\includegraphics[bb=50 240 580 600,width=17.cm]{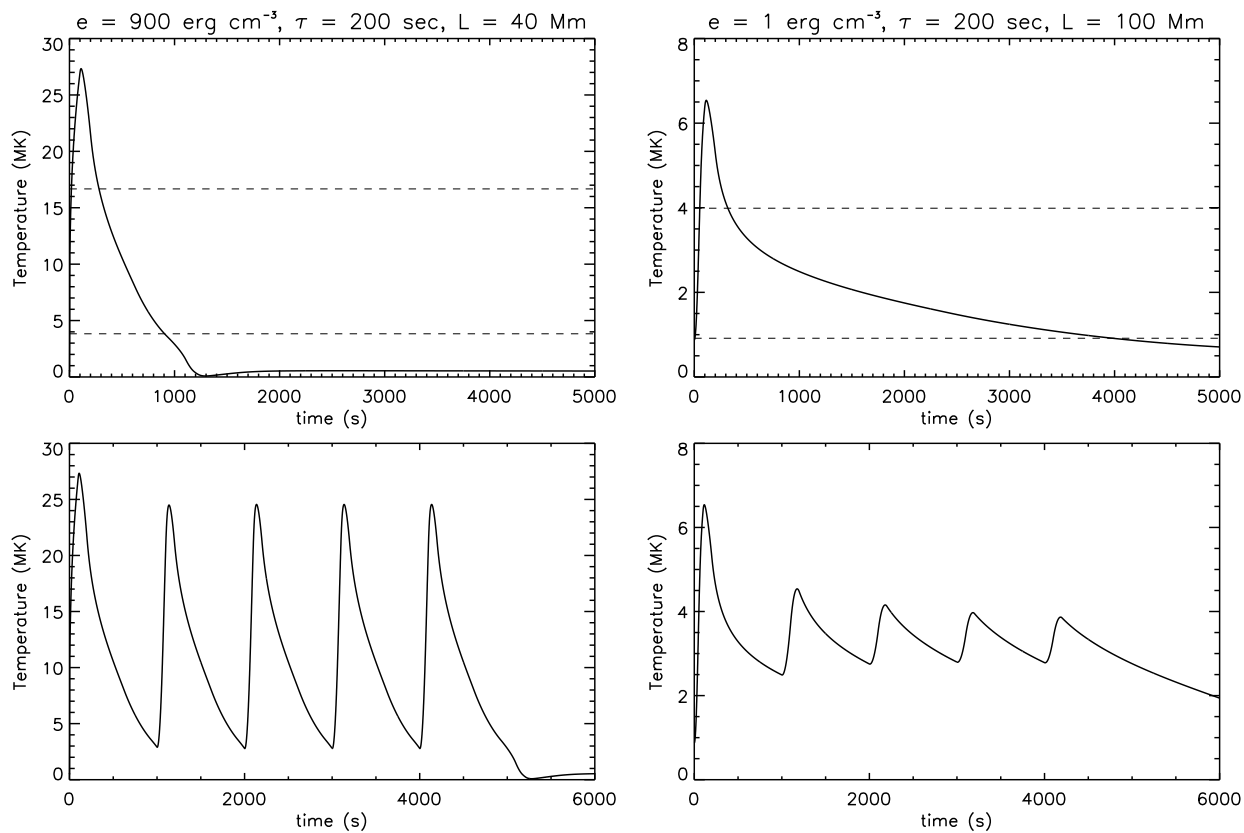}
       \caption{Temperature evolution for different nanoflare energies and loop lengths. Energy ($e$), loop length ($L$) and nanoflare durations ($\tau$) are provided on top of the upper panels. Upper panels: single nanoflares. The dashed horizontal lines indicate the 61\% and 14\% of the maximum temperature levels. Lower panels: evolution for 5 identical nanoflares separated by 1000 s times.}
\label{temp_evol}
\end{figure*}

We start by defining three categories of nanoflare frequency:

\begin{eqnarray*}
\textrm{High frequency:} & t_{rep}  <  0.5~t_{cool}, \\
\textrm{Intermediate frequency:} & 0.5~t_{cool}  <  t_{rep}  <  2~t_{cool}, \\
\textrm{Low frequency:} & t_{rep}  >  2~t_{cool},
\end{eqnarray*}

\noindent where $t_{rep}$ is the repetition time between consecutive events and $t_{cool}$ is the cooling time. The intermediate frequency interval has not been discussed much in the past, but there is a new appreciation that many nanoflares may be of this type (Cargill 2014). It is convenient to define the frequency categories in terms of the temperature change of the cooling plasma. The top panels of Figure~\ref{temp_evol}, explained below, show that the temperature evolution after the nanoflare can be approximated by an exponential:  $T(t) = T_0~\exp(-t/t_{cool})$, where $T_0$ is the maximum temperature reached during the event (see also Terzo et al. 2011). The temperature ratios $T(t_{rep})/T_0$ for each of the above frequency categories are then

\begin{eqnarray*}
\textrm{High frequency:} & T(t_{rep})/T_0 > 0.61, \\
\textrm{Intermediate frequency:} & 0.14 <  T(t_{rep})/T_0  <  0.61, \\
\textrm{Low frequency:} & T(t_{rep})/T_0 < 0.14.
\end{eqnarray*}

\noindent We determine statistics on nanoflare frequency in our models using these temperature ratios.\\

 The repetition times in our models are usually close to the timestep of the driving phase, 1000 s, the time it takes to displace a footpoint and increase the stress variable $S$ by an amount $d$ = 1000 km. The reason for this is that most strands are close to the critical condition, so they are pushed over the limit during most timesteps. The strands are near critical because the destressing that occurs at the end of the timestep is also $d$. This causes $S$ to fluctuate above and below the critical value. It is relatively uncommon for a strand to experience many events at one time and to relax far below the critical condition, i.e., ''avalanches'' are rare. The fact that strands fluctuate close to the critical condition also explains why the range of nanoflare energies is relatively narrow (Figure~\ref{power-law} and Table~\ref{parameters2}).

Table~\ref{freq_length} indicates the relative occurrence of low, intermediate, and high frequency nanoflares as a function of loop length. We have assumed a nanoflare duration of 200 s and used Equation~\ref{b_lock} to relate the magnetic field strength to loop length. We see that low-frequency nanoflares are more common in short loops, and high-frequency nanoflares are more common in long loops. This is a direct consequence of the dependence of the cooling time on temperature, since the repetition time is similar for all models, as discussed above. The initial cooling after a nanoflare is dominated by thermal conduction, for which the cooling timescale is proportional to $L^2 T^{-5/2}$ (L\'opez Fuentes, Klimchuk \& Mandrini 2007). Short loops cool more quickly because of the length dependence, and also because they have a greater field strength and more energetic nanoflares, leading to higher temperatures. More rapid cooling at a fixed repetition time means lower frequency nanoflares as defined by $T(t_{rep})/T_0$. Note that by combining Equations~\ref{e_scaling} and \ref{b_lock}, the nanoflare energy scales as $L^{-2.76}$, so the dependence of frequency on loop length is strong, as seen in the table.

To illustrate this point, we examine two cases:  a high energy nanoflare ($e =$ 900 erg cm$^{-3}$) in a short loop ($L =$ 40 Mm), and a low energy nanoflare ($e =$ 1 erg cm$^{-3}$) in a long loop ($L =$ 100 Mm). The nanoflare duration is 200 s. The $B$-$L$ relationship in these cases is stronger than given by Equations~\ref{b_lock}, so they should be considered rather extreme examples. The upper panels of Figure~\ref{temp_evol} show the temperature evolution. For guidance, the dashed horizontal lines in the plots correspond to the 0.61 $T_0$ and 0.14 $T_0$ levels. It can be easily seen that in the high energy event (upper-left panel) the temperature falls off very rapidly, so that the cooling time is relatively short. If a second nanoflare were to follow 1000 s later, the temperature would have decreased below the 14\% level relative to the maximum temperature. The situation is very different for the low energy case of the upper-right panel, in which the temperature reaches the "low frequency level" roughly 4000 s after the nanoflare started.

The difference between the cases is further illustrated in the lower panels of Figure~\ref{temp_evol}. There, we show the same nanoflares as in the upper panels but repeating at a 1000 s rate, corresponding to the cellular automaton time step duration. This is the minimum time delay between consecutive nanoflares in our models, and comparable to the typical delay, as discussed above. Clearly, according to our frequency interval definitions, we would classify nanoflares in the lower-left panel as low frequency and those in the lower-right panel as high frequency. Note that the initial static equilibrium conditions in the loop at $t = 0$ s are different from the conditions at start of the subsequent nanoflares. It takes several cycles before knowledge of the initial conditions is lost, especially in the example of the lower-right.

Table~\ref{freq_tau} indicates how the nanoflare frequency as defined by $T(t_{rep})/T_0$ depends on the nanoflare duration $\tau$. The runs used $L=80$ Mm and $B=211$ G. As expected, the percentage of high-frequency events increases with nanoflare duration. The longer the heating is switched on, the less time there is for the plasma to cool before the next event. The percentage of low-frequency nanoflares is small at all $\tau$ for this combination of $L$ and $B$. Smaller $L$ and/or larger $B$ would produce stronger nanoflares and therefore shorter cooling times and a higher percentage of low-frequency events.

%--------------CONCLUSIONS----------------------------------------------

\section{Discussion and conclusions}
\label{conclusions}

We study the problem of nanoflare heating of coronal loops using a 2D cellular automaton (CA) model based on Parker's (1988) idea of footpoint shuffling and tangling of elementary magnetic strands. To determine the plasma response to the heating we use the Enthalpy Based Thermal Evolution of Loops (EBTEL) model. From the computed temperature and density evolutions and the known response of coronal observing instruments we simulate observed lightcurves. We study the dependence of the model's output on the relevant physical parameters and we find and analyze a series of predicted scalings that can be compared with future observations.

Two primary results of our study concern the number distribution of nanoflares as a function of energy and the frequency with which nanoflares repeat on a given strand. We find that the number distribution obeys a power law with a slope of approximately -2.5. This is a robust result, with a standard deviation of 17\% as we vary the model parameters. For many years researchers have extrapolated power laws measured for flares and other resolvable events to lower energies in order to determine whether the corona could be heated by nanoflares. As pointed out by Hudson (1991), the slope of the distribution must be steeper than -2 in order for nanoflares to be energetically important. The results have been mixed, ranging from less steep (Aschwanden \& Parnell 2002 and references therein) to more steep (Krucker \& Benz 1998; Parnell and Jupp 2000; Benz 2004; Pauluhn \& Solanki 2007; Bazarghan et al. 2008), with shallower slopes tending to come from studies of flares and steeper slopes tending to come from studies of smaller impulsive events. The variation reflects the difficulty in measuring the slope, in part because of the assumptions that must be made in estimating the total energy that is released. We would also point out that there is no compelling reason to believe that the slope should be constant over the full range extending from large flares to nanoflares.

From an observational perspective, the nanoflare repetition frequency is important only insofar as the delay between successive events is longer than, shorter than, or comparable to the plasma cooling time. In our models, the nanoflare frequency increases with loop length. Since hot (X-ray) loops are best explained by high-frequency nanoflares, and warm (EUV) loops are best explained by low-frequency nanoflares, this would suggest that hot loops should be longer than warm loops. This is not observed to be the case, however. Hot loops are more prevalent in the cores of active regions, while warm loops are more prevalent outside the core. Most core emission is contained in a diffuse component, however, rather than in observationally distinct loops (Viall \& Klimchuk 2011). Whether this diffuse emission is better explained by high or low frequency nanoflares is a matter of debate. See the studies summarized in Table 3 of Bradshaw et al. (2012) and the discussion of uncertainties in Guennou et al. (2013). Subramanian et al. (2014) have examined diffuse emission outside the core, and again the results are inconclusive regarding nanoflare frequency. Cargill (2014) recently suggested that all the results might be reconciled if nanoflares have random frequencies centered about a mean value that is comparable to a cooling time, i.e., intermediate frequency, and if there is a relationship between the event size and delay between events.

The predictions of our model of course depend upon the assumptions that go into it. We have three main assumptions. First, magnetic reconnection does not occur unless a critical condition corresponding the misalignment angle between adjacent magnetic strands is met. This is supported by theoretical studies of the secondary instability of current sheets (Dahlburg et al. 2005, 2009). Second, the test for criticality does not happen until after a full stressing step has been made. In other words, reconnection holds off at least until photospheric convection has transported the magnetic footpoint a characteristic distance $d = 1$ Mm. The critical condition can therefore be exceeded, though not by a large amount. Third, reconnection, once it occurs, releases an amount of stress equal to the stress added during a driving step. It does not cause the total stress to drop far below the critical value. A strand can reconnect with multiple strands during a single iteration, but in practice the number rarely exceeds three. The mean is close to one. As a result the stress tends to hover around the critical value. The energy of the nanoflare, which combines the energies of the separate reconnections, does not have a wide variation. The energy range is typically less than 1.0 in the logarithm, i.e., factor of 10 difference between the largest and smallest for a given set of model parameters.

Whether these assumptions are reasonable has yet to be determined. The detailed physical scenario of nanoflares is still unknown. We note that  Ugarte-Urra \& Warren (2014) used a combination of observations and hydro modeling to infer a nanoflare repetition rate of 2 to 3 per hour. This corresponds to a delay of  $\sim$1400 s, which agrees well with our assumed delay of 1000 s. From another observational/modeling comparison, Cargill (2014) suggests a delay of a few hundred to somewhat over 2000 s. Finally, Dahlburg et al. (2005) suggest approximately 2000 s based on the MHD simulations of the secondary instability. It is clear that much more work is left to be done.

\acknowledgements The authors acknowledge useful comments and suggestions from the anonymous referee. They also wish to thank Dr. Peter Cargill for useful discussions. JAK's work was funded by the NASA Supporting Research and Technology and Guest Investigator programs. MLF acknowledges financial support from the Argentinean grants PICT 2012-0973 (ANPCyT), UBACyT 20020100100733 and PIP 2009-100766 (CONICET).

%----------------------REFERENCES---------------------------------------

%-------------TABLES----------------------------------------

\clearpage
\begin{table}
\caption{Numerical values used for the parameters of the model.}
\vspace{0.5cm}
\label{parameters}
$\begin{array}{lcccccc}
\hline
B_{v} $(Gauss)$   & 40   & 100  & \bf{200}  & 500 \\
L $(Mm)$          & 40   &  60  & \bf{80}   & 100 & 120 \\
\tan \theta_{c}   & 0.15 & 0.2  & \bf{0.25} & 0.3 & 0.35 \\
N                 & 9    & 25   & \bf{49}   & 81  & 121  & 169 \\
\tau 		  & 50   & 100  & \bf{200}  & 500 & 1000 \\	
\hline
\end{array}$
\end{table}

\begin{table}
\caption{Power-law indices and nanoflare energy ranges.}
\vspace{0.5cm}
\label{parameters2}
$\begin{array}{cccc}
\vspace{0.1cm}
L($Mm$) & B($G$) & \Delta $E range($10^{24}$erg)$ & $Power-law index$ \\
\hline
40   &  389  &  557-2465  &	-2.43 \pm 0.14 \\
60   &  272  &  226-722   &	-2.40 \pm 0.19 \\
80   &	211  &	139-394   &	-2.73 \pm 0.14 \\	
100  &	174  &	101-290   &	-2.64 \pm 0.11 \\	
120  & 	148  &   70-174   &	-2.32 \pm 0.14 \\
\hline
\end{array}$
\end{table}

\begin{table}
\caption{Distribution of nanoflare frequency versus strand length for the intervals defined in Section~\ref{frequency}.}
\vspace{0.5cm}
\label{freq_length}
$\begin{array}{cccc}
\vspace{0.1cm}
L($Mm$) & $Low (\%)$ & $Intermediate (\%)$ & $High (\%)$ \\
\hline
40   & 43  &  52    &  5	\\
60   & 33  &  51    &  16 	\\
80   & 9   &  48    &  43	\\	
100  & 7   &  50    &  43	\\	
120  & 1   &  45    &  54	\\
\hline
\end{array}$
\end{table}

\begin{table}
\caption{Idem Table~\ref{freq_length} for $\tau$ dependence.}
\vspace{0.5cm}
\label{freq_tau}
$\begin{array}{cccc}
\vspace{0.1cm}
\tau($sec$) & $Low (\%)$ & $Intermediate (\%)$ & $High (\%)$ \\
\hline
200   & 7  &  61  &  32	\\
400   & 7  &  45  &  48 \\
600   & 5  &  42  &  53	\\	
800   & 6  &  37  &  57	\\	
1000  & 8  &  34  &  58	\\
\hline
\end{array}$
\end{table}

%---------FIGURES--------------------------------------------

\clearpage

\clearpage

\clearpage

\clearpage

\clearpage

\clearpage

\clearpage

\clearpage

\end{document}